\documentclass[twocolumn]{article}

\usepackage[margin=1in]{geometry}

\usepackage{times}
\usepackage{graphicx}
\usepackage{amsmath}
\usepackage{microtype}
\usepackage[hyphens]{url}
\usepackage{hyperref}
\usepackage{natbib}
\usepackage{caption}

\usepackage{algorithm}
\usepackage{algorithmic}
\usepackage{multirow}

\usepackage{listings}
\title{Broken-Token: Filtering Obfuscated Prompts by Counting Characters-Per-Token}

\author{
    Shaked Zychlinski$^*$ \qquad Yuval Kainan$^*$ \\[0.5em]
    \textit{JFrog} \\
    \texttt{\{shakedz, yuvalk\}@jfrog.com} \\[0.5em]
    {\small $^*$Equal contribution}
}

\date{}
\begin{document}

\twocolumn[
\maketitle
\vspace{-3em}
{\small
\begin{quotation}
\begin{abstract}
	\hspace*{\parindent}	
	Large Language Models (LLMs) are susceptible to jailbreak attacks where malicious prompts are disguised using ciphers and character-level 
	encodings to bypass safety guardrails. While these guardrails often fail to interpret the encoded content, the underlying models can still process 
	the harmful instructions. We introduce \textbf{CPT-Filtering}, a novel, model-agnostic with negligible-costs and \textbf{near-perfect accuracy} 
	guardrail technique that aims to mitigate these attacks by 
	leveraging the intrinsic behavior of Byte-Pair Encoding (BPE) tokenizers. Our method is based on the principle that tokenizers, trained on natural 
	language, represent out-of-distribution text, such as ciphers, using a significantly higher number of shorter tokens.

	Our technique uses a simple yet powerful artifact of using language models: the average number of Characters Per Token (CPT) in the text. 
    This approach is motivated by the high compute cost of modern methods - relying on added modules such as dedicated LLMs or perplexity models.
    We validate our approach across a large dataset of over 100,000 prompts, testing numerous encoding schemes with several popular tokenizers. 
    Our experiments demonstrate that a simple CPT threshold robustly identifies encoded text with high accuracy, even for very short inputs. 
    CPT-Filtering provides a practical defense layer that can be immediately deployed for real-time text filtering and offline data curation.
\end{abstract}

\end{quotation}
}
\vspace{1em}
]

{\raggedright\noindent\textbf{Dataset:} \url{https://huggingface.co/datasets/jfrog/obfuscation-identification}\par}

\begin{figure}[!htb]
    \centering
    \includegraphics[width=0.98\columnwidth]{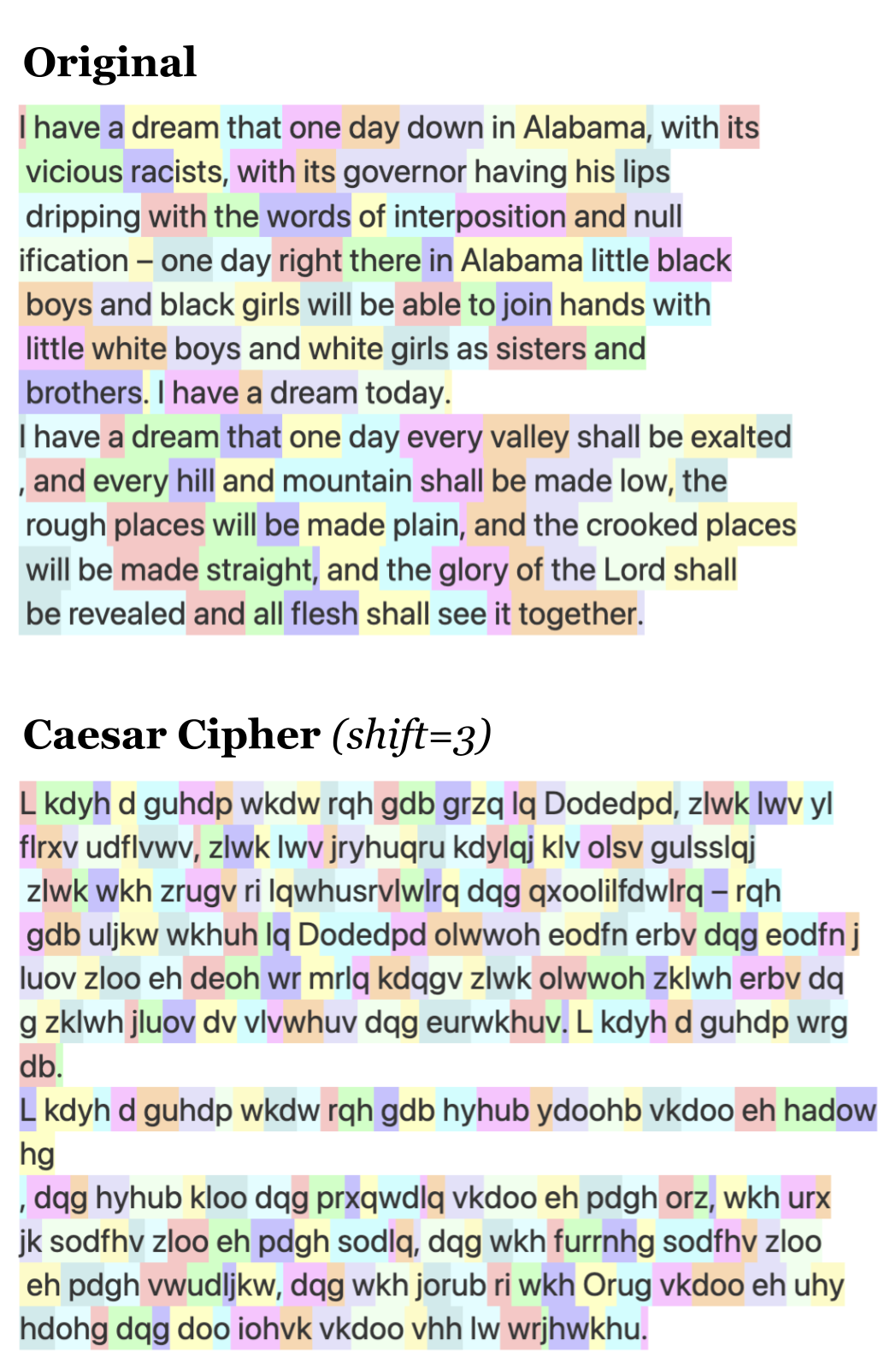} 
    \caption{Example of a tokenized representation (using GPT-4o) of an English text (top) and a ciphered version of it using Caesar Cipher (bottom). 
	         While both containing the same number of characters (613), the original version is constructed of 128 tokens, while the ciphered version 
			 is constructed of 294 tokens --- more than double.}
    \label{fig:example}
\end{figure}

\begin{table*}[htb]
    \centering
    \begin{tabular}{l l r}
        \hline
        \textbf{Hugging Face dataset name} & \textbf{Column Used} & \textbf{Dataset size} \\
        \hline
        SoftAge-AI\slash prompt-eng\_dataset & Prompt & 1k \\
        Aiden07\slash dota2\_instruct\_prompt & instruction & 4.74k \\
        hassanjbara\slash ghostbuster-prompts & text & 2.18k \\
        k-mktr\slash llm\_eval\_prompts & prompt & 2.12k \\
        facebook\slash cyberseceval3-visual-prompt-injection & user\_input\_text & 1k \\
        HacksHaven\slash science-on-a-sphere-prompt-completions & prompt & 7.08k \\
        grossjct\slash ethical\_decision\_making\_prompts & prompt & 1.76k \\
        \hline
    \end{tabular}
    \caption{Datasets used for evaluation. Our dataset was created by applying 5 text encoding techniques 
        (\textit{Caesar Cipher, Leetspeak, Reverse, Binary Encoding and Base64 Encoding}) to each prompt in the aforementioned datasets. 
        In each row, size is the number of rows in each dataset.}
    \label{tab:datasets}
\end{table*}

\begin{figure*}[tb]
    \centering
    \includegraphics[width=0.95\textwidth]{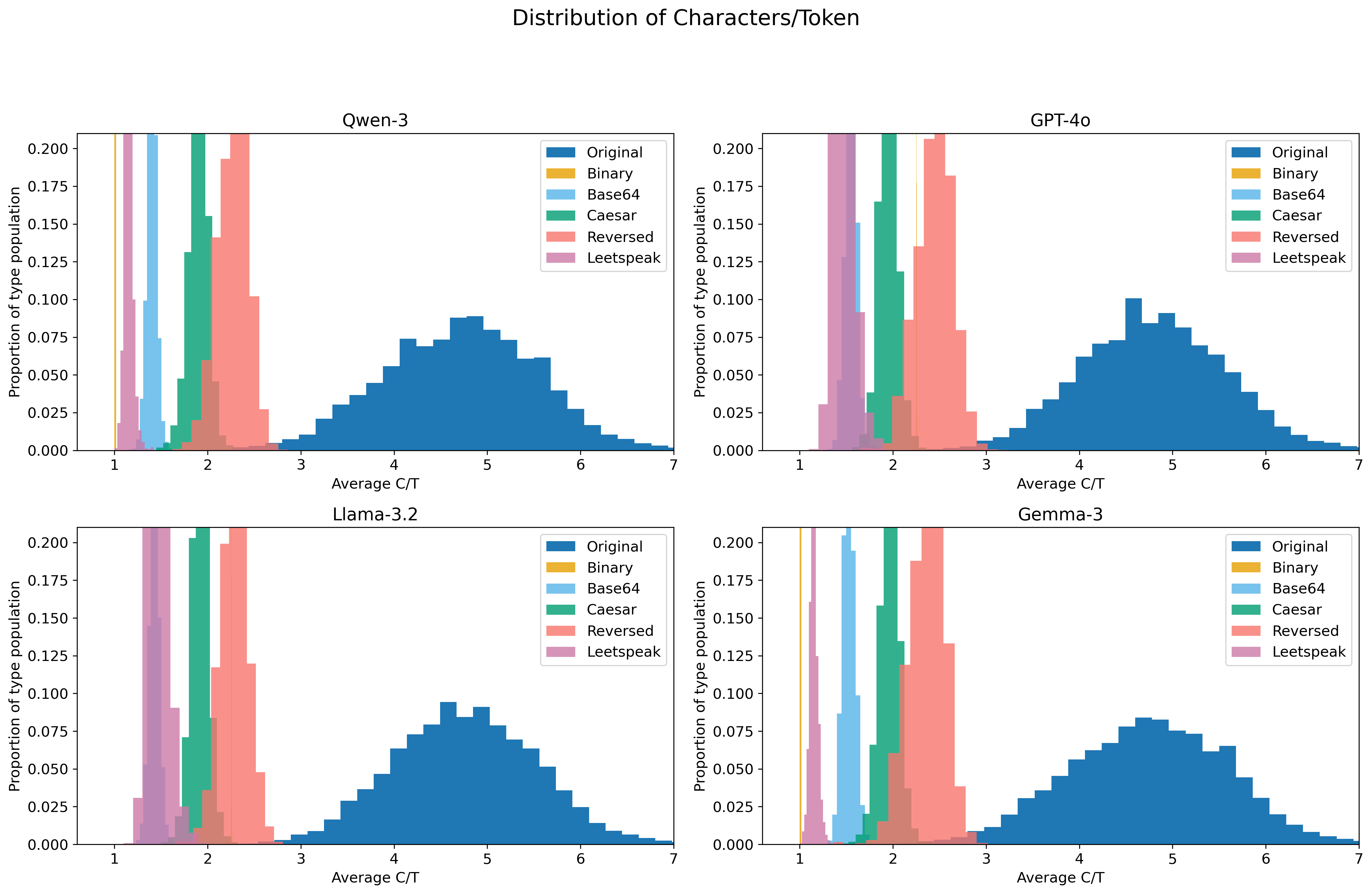} 
    \caption{The average number of characters-per-token for the four models checked (We cut the y-axis at $ 0.2 $ to better visualize the distributions)}
    \label{fig:dists}
\end{figure*}

\begin{table}[tb]
    \centering
    \begin{tabular}{l c c}
        \hline
        \textbf{Tokenizer} & \textbf{t-statistic} (all) & \textbf{p-value} (all)\\
        \hline
        Qwen-3 & 520 & $< 10^{-5}$ \\
        GPT-4o & 504 & $< 10^{-5}$ \\
		Llama-3.2 & 513 & $< 10^{-5}$ \\
        Gemma-3 & 499 & $< 10^{-5}$ \\
        \hline
    \end{tabular}
    \caption{t-test of the separation between the original and all obfuscated prompts (all other types combined) for each tokenizer.
	        (n\textsubscript{original}=19,816, n\textsubscript{obfuscated}=101,844. Using Welch's t-test for unequal variance).}
    \label{tab:t_test}
\end{table}

\begin{table*}[tb]
    \centering
    \renewcommand{\arraystretch}{1.3}
    \begin{tabular}{l l c c c c }
        \hline
        & & \textbf{Qwen 3} & \textbf{GPT-4o} & \textbf{Llama 3.2} & \textbf{Gemma 3}\\
        \hline
        \multicolumn{2}{c}{Average CPT Threshold} & 2.69 & 2.92 & 2.71 & 2.81 \\
        \hline 
        \multirow{2}{*}{Reversed} & Accuracy & 99.4\% & 99.4\% & 99.7\% & 99.1\% \\
        & F1 & 0.994 & 0.994 & 0.997 & 0.991 \\
        \multirow{2}{*}{Base64} & Accuracy & 99.6\% & 99.6\% & 99.8\% & 99.4\% \\
        & F1 & 0.996 & 0.996 & 0.998 & 0.994 \\
        \multirow{2}{*}{Caesar Cipher} & Accuracy & 99.6\% & 99.6\% & 99.8\% & 99.4\% \\
        & F1 & 0.996 & 0.996 & 0.998 & 0.994 \\
        \multirow{2}{*}{Leetspeak} & Accuracy & 99.6\% & 99.6\% & 99.8\% & 99.4\% \\
        & F1 & 0.996 & 0.996 & 0.998 & 0.994 \\
        \multirow{2}{*}{Binary} & Accuracy & 99.6\% & 99.6\% & 99.8\% & 99.4\% \\
        & F1 & 0.996 & 0.996 & 0.998 & 0.994 \\
        \hline 
        \multirow{2}{*}{\textbf{All}} & \textbf{Accuracy} & 99.8\% & 99.8\% & 99.9\% & 99.7\% \\
        & \textbf{F1} & 0.994 & 0.994 & 0.997 & 0.991 \\
        \hline
    \end{tabular}
    \caption{Optimal average CPT thresholds for examined tokenizers, along with the accuracy and F1 scores of distinguishing 
             obfuscated texts from regular ones.}
    \label{tab:thresholds}
\end{table*}

\begin{figure*}[tb]
    \centering
    \includegraphics[width=0.95\textwidth]{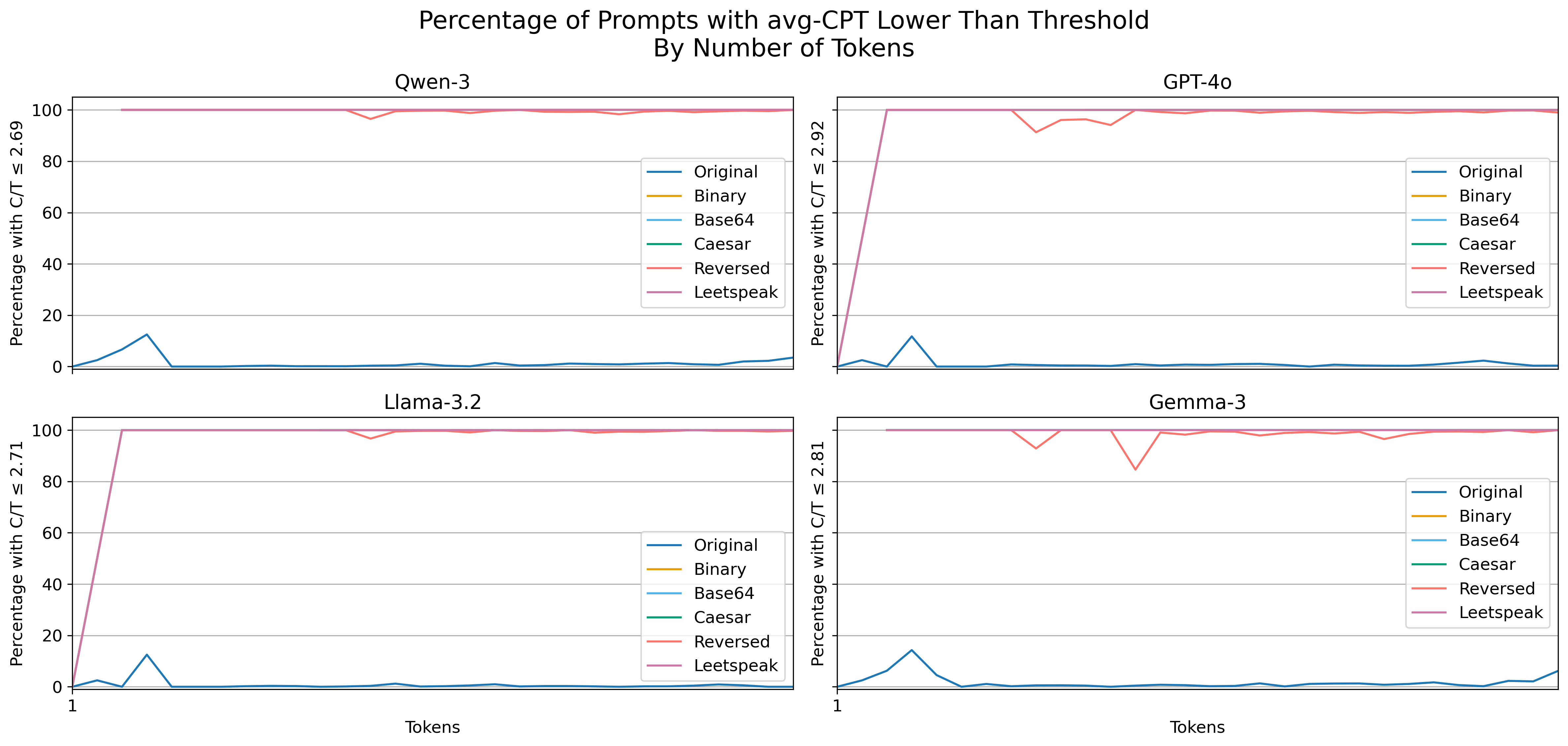} 
    \caption{\raggedright Percentage of prompts marked as ``obfuscated-prompts'' in each category, as a function of prompt length tokens}
    \label{fig:short_tokens}
\end{figure*}

\begin{figure}[tb]
    \centering
    \includegraphics[width=0.95\columnwidth]{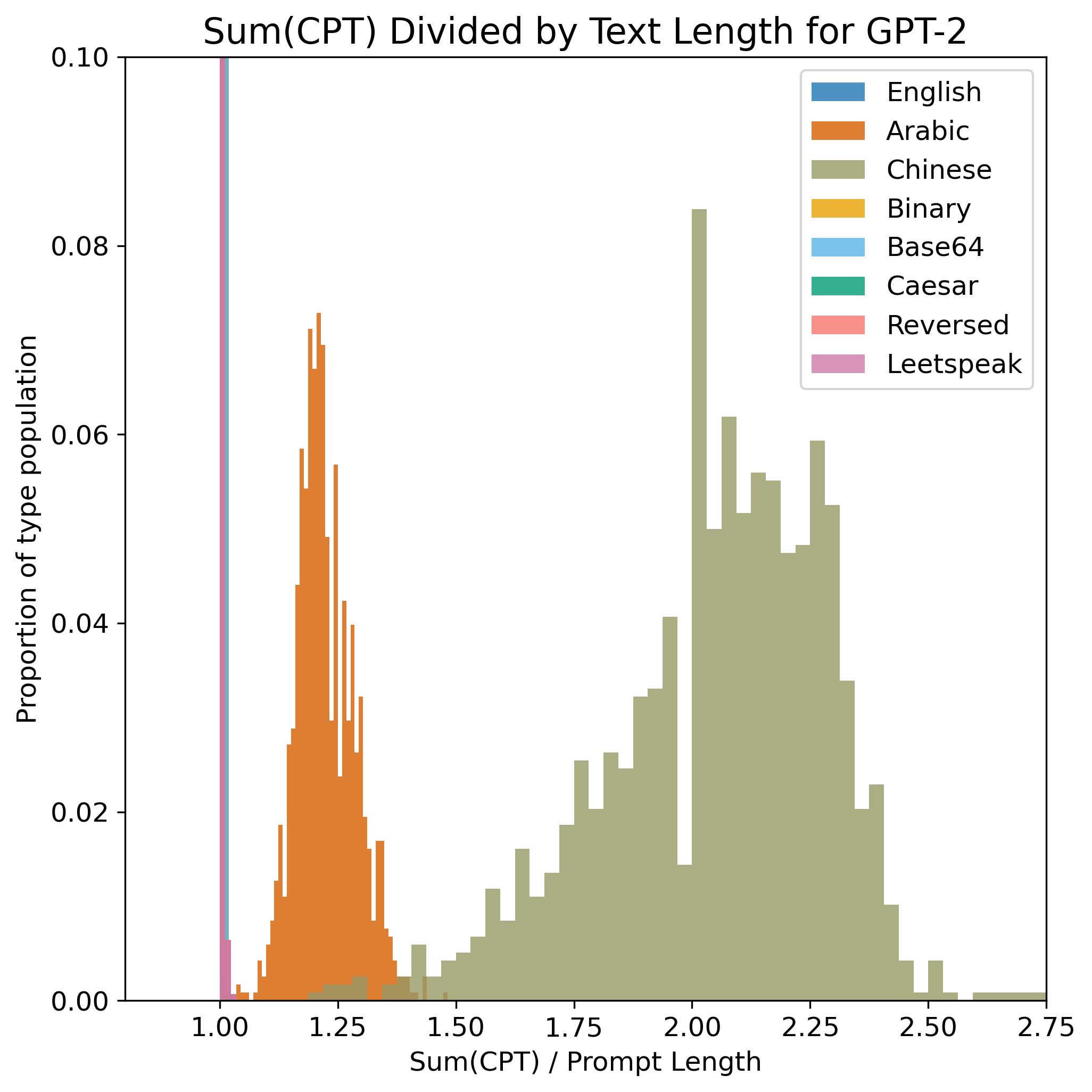}  
    \caption{\raggedright Sum of number of characters per token of a given prompt divided by the length (in characters) of the prompt, using GPT-2 tokenizer. 
                Other than Arabic and Chinese, all other classes are confined to 1.0}
    \label{fig:gpt2_sum}
\end{figure}

\begin{figure*}[tb]
    \centering
    \includegraphics[width=0.99\textwidth]{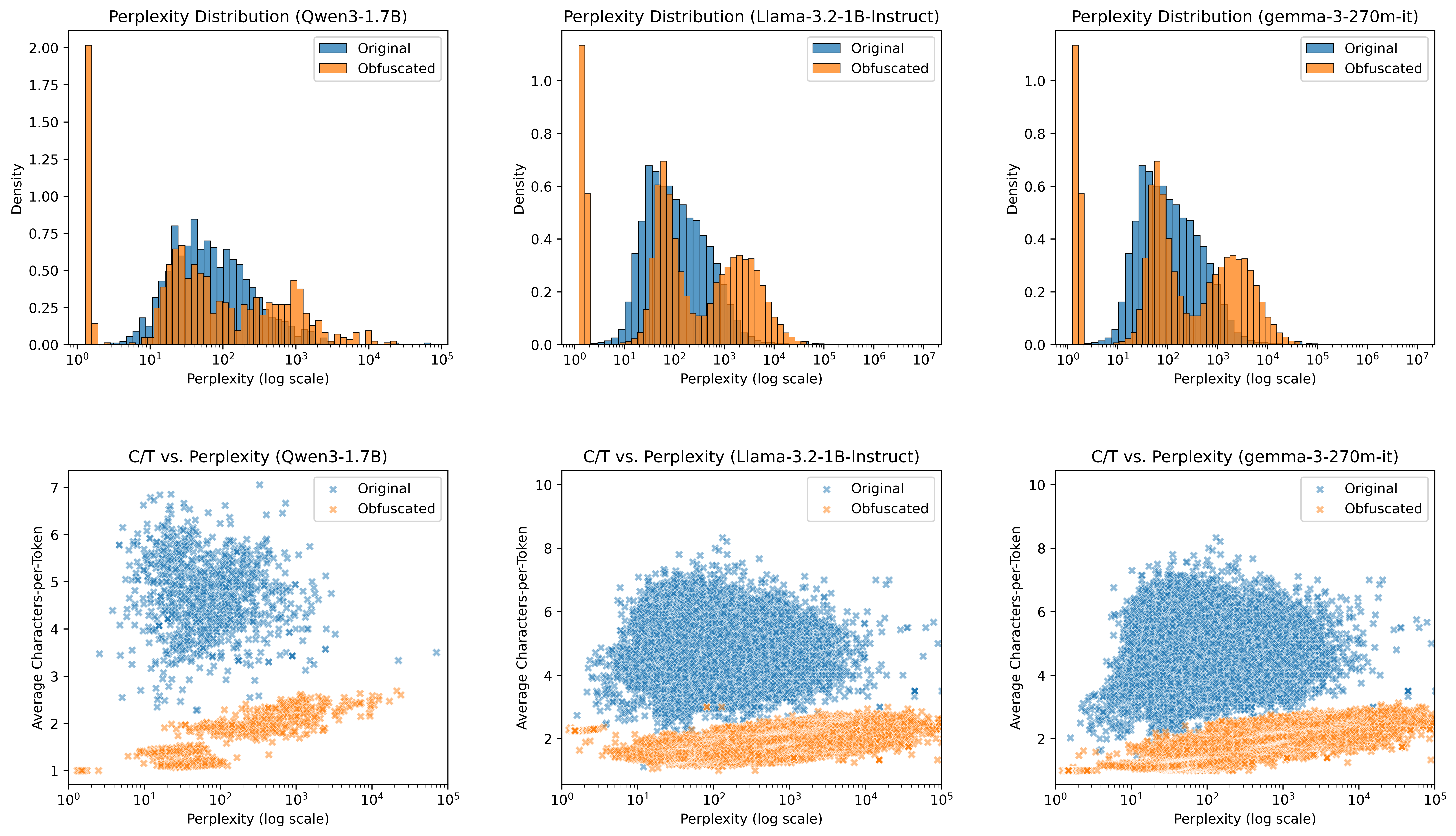} 
    \caption{\textit{Top:} the distribution of perplexity for the original and obfuscated prompts.
             \textit{Bottom:} the distribution of perplexity and average character-per-token for the original and obfuscated prompts.}
    \label{fig:ppl_cpt}
\end{figure*}

\begin{figure}[tb]
    \centering
    \includegraphics[width=0.9\columnwidth]{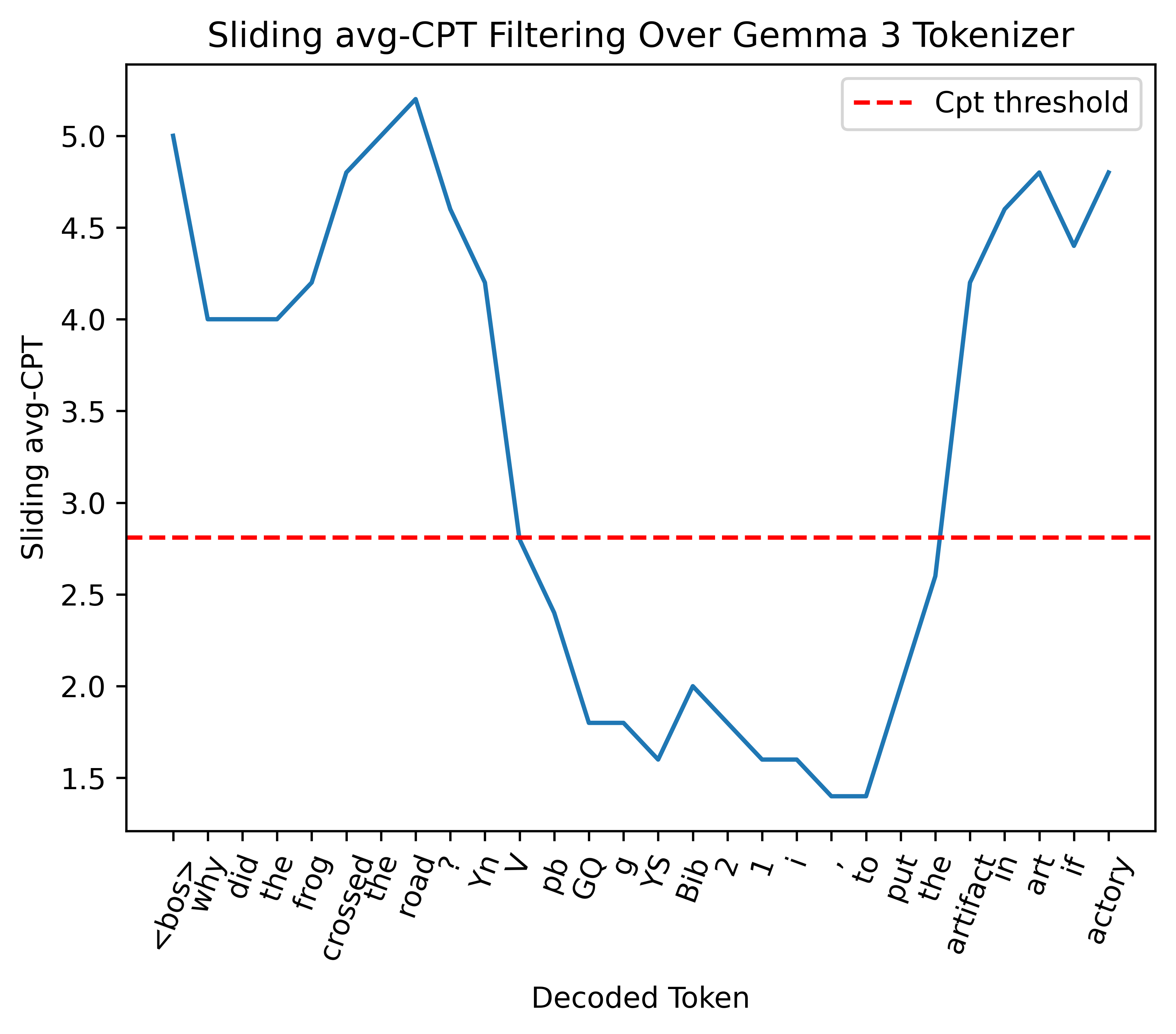} 
    \caption{\raggedright An example of applying sliding window CPT filtering, where the obfuscated text is mixed within the benign text. 
                Window size of 5 tokens, tokenizer used is Gemma-3. The red line is the model's base CPT threshold for Base64.}
    \label{fig:rolling_cpt}
\end{figure}

\section{Introduction}
A critical security flaw in modern Large Language Models (LLMs) allows malicious actors to bypass safety guardrails using prompts disguised with ciphers or other 
encodings \cite{12:survey, 16:baselinedefenses}. While these guardrails often fail to recognize the threat, the model itself can still interpret the harmful instructions, resulting in unaligned behavior.
We propose a novel detection method that turns a tokenizer into a defense mechanism using the computational cost of the tokenization process, with minimal additional cost. 
We exploit the fact that tokenizers (specifically BPE, see subsection on tokenizers) represent these out-of-distribution, obfuscated inputs inefficiently — using a much larger number of shorter tokens than for regular text. 
This paper demonstrates how measuring this tokenization anomaly provides a simple and highly effective method for identifying and blocking such attacks. 

\subsection{Tokenizers}
\label{sec:tokenizers}
At the core of how Large Language Models (LLMs) process and understand human language lies the tokenizer. 
A tokenizer's primary function is to break down a given text input into a sequence of smaller units, or "tokens," 
that the model can comprehend and process. One of the most prevalent tokenization strategies is Byte-Pair Encoding (BPE).  b
BPE operates by iteratively merging the most frequent pairs of bytes in a training corpus, building a vocabulary of substring units \cite{1:rarewords}, 
which can be as long as complete words. This allows the model to handle a vast vocabulary, including rare and out-of-vocabulary terms, 
by representing them as sequences of more common sub-terms. The way a prompt is tokenized directly impacts how the LLM interprets it, 
and an unexpected or unusual tokenization can lead to unpredictable model behavior \cite{3:badcharacters}. 

\subsection{LLM Jailbreaks}
The input text/prompt represents a significant attack surface for LLM security \cite{12:survey}. Prompt attacks, commonly known as "Jailbreaks", consist of maliciously crafted 
inputs designed to make an LLM execute unintended instructions or bypass its safety constraints \cite{5:autodan, 6:jailbroken}. Common techniques include instructing the model 
to adopt a persona without ethical restrictions (e.g., the 'Do Anything Now' or DAN prompt) \cite{5:autodan, 8:danreddit} or framing the malicious request as a purely 
hypothetical scenario for a fictional story \cite{9:ignoreprevious} and structural manipulations \cite{2:structuralsleight} .

As the technology matures, LLMs understand multiple languages, different modalities (visual data, audio, etc.), and obfuscated messages. With each new capability, new attack surfaces arise \cite{18:visual}. 
These jailbreaks (attack vectors) usually mean embedding harmful instructions in new types of input data, in various ways.
One notable embedding attack technique is obfuscation of text using ciphers and encoders, referred as Character Injection \cite{3:badcharacters, 12:survey}.
This approach is effective because LLMs are often capable of interpreting encoded or modified text while the defense system fails to handle it properly \cite{3:badcharacters,4:toosmart}. 
This discrepancy can arise because the defense system may have been trained on a different dataset than the underlying LLM, creating a key vulnerability \cite{3:badcharacters, 4:toosmart}. 
Research shows these techniques can be highly effective, in some cases achieving up to 100\% evasion success against prominent guardrail systems \cite{3:badcharacters, 4:toosmart, 12:survey}.

\subsection{Character-Based Manipulations}
Character-based manipulations involve altering the input text at the character level to obfuscate malicious intent from the LLM's safety filters while preserving the semantic meaning for the model 
to process. These techniques can be broadly categorized as follows:

\subsubsection{Common Obfuscation Techniques}
These methods utilize well-established encoding schemes to disguise harmful prompts \cite{3:badcharacters, 4:toosmart, 12:survey}. 
A comprehensive survey of these techniques can be found in \cite{3:badcharacters, 12:survey}. Commonly used encodings methods include:
\begin{itemize}
    \item \textbf{Binary Encoding}
    \item \textbf{Base64 Encoding}
    \item \textbf{Caesar Cipher:} A simple substitution cipher where each letter in the plaintext is shifted a certain number of places down or up the alphabet.
    \item \textbf{Reversed Text:} A straightforward technique where the malicious prompt is simply written backwards.
    \item \textbf{Leetspeak:} Also known as ``1337'', this method replaces standard letters with numbers or symbols that visually resemble them (e.g., ``A'' becomes ``4'', ``E'' becomes ``3'').
\end{itemize}

\subsubsection{Custom Obfuscation Techniques}
More sophisticated attacks employ novel and complex ciphers specifically designed to evade LLM guardrails.
\begin{itemize}
    \item \textbf{Novel Complex Ciphers:} Research such as \cite{4:toosmart, 7:llamaspace, 14:complex} has demonstrated the development and effectiveness of unique, multi-layered encryption methods created specifically for jailbreak purposes.
    \item \textbf{Uncommon Text-Encoded Structures:} Another avenue of research, detailed in \cite{2:structuralsleight}, explores the use of obscure or non-standard text encodings alongside structural manipulation to bypass detection mechanisms.
\end{itemize}

\subsection{Contributions}
We introduce a new guardrail method, \textbf{CPT-Filtering} (Characters-Per-Token Filtering), that allows us to perform \textbf{out-of-distribution detection, with no extra cost}, leveraging the fact that one always tokenizes the input text before passing the tokens to the LLMs.
Our method could be used as a defense mechanism (guardrail) in various ways, two prominent ones are:

\begin{enumerate}
    \item In inference-time, our method can be used as a guardrail for filtering encoded and/or ciphered prompts, which tend to be used for jailbreak purposes.
    \item In training or fine-tuning time, this method can be used for data curation to mitigate data poisoning using encoded and ciphered texts and general non-natural text data points.
\end{enumerate}

We also present our dataset of 120k prompts, both obfuscated and natural, to support further research in the community.

\section{Related Works}

\subsection{LLM Guardrails}
To detect jailbreak attacks, multiple tools had been introduced in recent years \cite{12:survey, 16:baselinedefenses, 17:sok}. 
These tools vary in performance (computational cost, latency, accuracy), and implement multiple core algorithms, including perplexity 
filtering \cite{16:baselinedefenses} and implementing LLM-based systems \cite{12:survey, 17:sok}.

\subsection{Perplexity-Based Guardrails}
Classical approaches for character level guardrails is to use perplexity as the measurement of the ``normality'' of the prompt.
In \cite{16:baselinedefenses}, the authors present two approaches to use perplexity as a guardrail:
\begin{itemize}
    \item A simple filter that checks the overall perplexity of a prompt against a threshold.
    \item A windowed perplexity filter that analyzes contiguous chunks of text, allowing for sub-prompt level detection.
\end{itemize}

\subsection{LLM-Based Guardrails}
LLM-based classifiers can be used to screen prompts and responses. Many solutions exist in this domain, for example Llama Guard \cite{10:llamaguard}, an 8B parameters SLM-based guardrail model designed for Human-AI conversations.
In \cite{19:nemo}, the authors introduce "NeMo Programmable Guardrails" --- a toolkit for a custom guardrail system. NeMo is powered by an LLM that decomposes the users prompt to understand its intent and semantics for a KNN-FAISS based miss-use classifier.

\subsection{N-gram Perplexity-Based Guardrails}
A more nuanced approach is to use a custom model for threat detection at the sub-prompt level. 
In \cite{15:ngram}, the authors had trained an N-gram perplexity model (a look up table), used to measure the perplexity of the N-grams decomposition of the prompt. 
This sequence is then used to approximate the threat score by measuring the ``text fluency'' of the prompt.
\section{Method}
\raggedright We propose a new guardrail technique, named \textbf{CPT-Filtering}, drastically reducing the risk of ciphered attacks. 
Our method is a low-compute, model-agnostic, outlier detection mechanism, detecting outlier prompts using their abnormal tokenized representation. 
We propose a new technique of assessing whether a text is encoded or encrypted based solely on a BPE-based tokenizer. 
The method requires no pre-training nor prior knowledge about the encoding technique. The core idea lies in the training process of BPE tokenizers: 
strings must appear sufficient times in the training data in order to be given a single token to represent them. 
While this is true for most terms, this does not apply to encoded or ciphered texts, forcing the tokenizer to construct them from short-string tokens. 
\textbf{In other words, encoded texts will be constructed of \textit{more} tokens and \textit{shorter} token-values (less characters per token) than the same unencoded text}
(see Figure~\ref{fig:example}).

\subsection{Dataset}
To verify our hypothesis, we created a dataset of approximately 20k English prompts from 7 different open datasets (see Table~\ref{tab:datasets}).
We then applied each of the following encoding methods to each prompt:
\begin{itemize}
    \item Binary Encoding (8-bit)
    \item Base64 Encoding
    \item Caesar Cipher (shift=3)
    \item Reversed Text
    \item Leetspeak (see Table~\ref{tab:leetspeak} in the Appendix)
\end{itemize}
The combination of all these methods along with the original prompts resulted in a dataset of approximately 120k prompts. 
We provide this dataset in the link found in the links section, as well as the code to recreate it in the Code Appendix. 
Figure~\ref{fig:dataset} in the appendix shows the distribution of the length of the prompts in the dataset. 

\subsection{Evaluating Average Characters-Per-Token}
We tested four common and widely used tokenizers, referenced to by their model names:
\begin{itemize}
    \item Qwen 3 (\textasciitilde151k tokens) \cite{qwen3}
    \item OpenAI GPT-4o (\textasciitilde200k tokens) \cite{gpt4o}
    \item Meta Llama 3.2 (\textasciitilde128k tokens) \cite{llama3_2}
    \item Google Gemma 3 (\textasciitilde262k tokens) \cite{gemma3}
\end{itemize}
Each tokenizer has a different number of known tokens, and was trained on different datasets - both effecting directly on the selection of substrings to tokenize.

Per each tokenizer, we calculated the average number of characters per token (a.k.a $ \overline{\mathrm{CPT}} $) for each prompt in the dataset.
Figure~\ref{fig:dists} presents the distributions we received. We clearly see a well separation between the distributions of the original prompts and their encoded variations.
T-Test of the separation between the original and obfuscated prompts (all other types combined) for each tokenizer is shown in Table~\ref{tab:t_test}.
All p-values were lower than $10^{-5}$. We also performed a t-test for each original-vs-obfuscated type separately, and all p-values were still lower than $10^{-5}$.
We present the results in Table~\ref{tab:t_test_full} in the appendix.
We also computed the optimal average-CPT threshold per model, using a precision-recall plot\footnote{We used \texttt{dython} for this computation \cite{dython}} 
over all obfuscated texts at once, and then computed the 
accuracy and F1 score of the separation of obfuscated texts from regular ones. Results are shown is Table \ref{tab:thresholds}.
As can be easily seen, results are \textbf{near-perfect} for all models and all obfuscation techniques examined.

\subsection{Prompt Length}
As our method uses the prompt construction by tokens for detection, we wish to verify what is the minimal length to which our method applies.
The length of a text can be measured in two methods: characters and tokens.
Figure~\ref{fig:short_tokens} shows the percentage of prompts marked as ``obfuscated-prompts'' (meaning, $\overline{\mathrm{CPT}} < $ threshold) in each category,
as a function of the prompt length in tokens (Figure~\ref{fig:short_chars} in the appendix displays the same for length in characters).
We clearly see our method can be used to detect encoded prompts even for very short texts (length of \textasciitilde3 tokens
for all examined models).

\subsection{Multilingual Scenarios}
When examining average characters-per-token for languages which do not rely on alphanumeric characters, we witness a sharp drop in the ability of 
$ \overline{\mathrm{CPT}} $ to distinguish between obfuscated prompts to languages such as Chinese and Arabic, as seen in Figure \ref{fig:langs_no_good} in
the appendix. To mitigate this issue, we add a second, \textit{monolingual}, tokenizer, such as GPT-2 \cite{gpt2}.

As GPT-2 was trained over only alphanumeric texts, the way it handles non-alphanumeric texts causes a unique effect, in which the tokenizer 
is usually required to use more than one token per each sign/letter in the text, where each token value by itself has nothing to do with the actual text. 
For example, the Chinese text \includegraphics[height=1em]{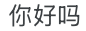}, which consists of three symbols, requires GPT-2 to 
use \textit{six} tokens. Each of these tokens, when decoded separately, results in a string of at least one character (and sometimes more).
In other words, for non-English prompts, the sum of the length of the values of each token separately will be larger than the length of the text itself:
$$ 
\begin{gathered}
\text{for token } t \text{ in prompt } p\!:\; \\
 \frac{\sum_{t}{\mathrm{len}(\mathrm{decode}(t))}}{\mathrm{len}(p)} \begin{cases}
    > 1 & \text{non-alphanumeric texts} \\
    = 1 & \text{alphanumeric texts}
\end{cases}
\end{gathered}
$$
This clear separation can be seen in Figure \ref{fig:gpt2_sum}. This allows us to distinguish non-alphanumeric prompts from alphanumeric,
and regular alphanumeric texts from alphanumeric obfuscated prompts (such as Base64, Binary, Leetspeak, etc). 
We refrain from discussing non-alphanumeric obfuscated prompts (i.e. reversed Chinese texts or Arabic Caesar Cipher) due to the lack
of concrete evidence of the success rates of such attacks.

\subsection{Comparing CPT-Filtering and Perplexity-Filtering}
We compare our CPT-Filtering with standard perplexity (PPL) filtering. We present here our measurements for three models: Qwen3-1.7B, Llama3.2-1B-it, and Gemma 3-270m-it. 
These models were chosen due to both their popularity and their size, making them viable for real-world use perplexity filtering.
We assign a label to each prompt to be either "Original" or "Obfuscated" (all encoded or ciphered prompts).
We then calculate the PPL and CPT per input. As seen in Figure~\ref{fig:ppl_cpt}, PPL-based analysis only does not provide a clear separation.
By adding the CPT-Filtering information, the separation becomes clear.

\subsection{Mixed Inputs}
So far, we discussed the case of fully obfuscated prompts only. Yet, obfuscation attacks might contain a mixed input - meaning, 
normal "safe" text with obfuscated part in it. Mixed texts will obviously have higher $\overline{\mathrm{CPT}} $ then obfuscated texts. 
As discussed in "Prompt Length" subsection earlier, CPT Filtering is effective even for very short inputs, enabling us to detect obfuscated parts 
of at least \textasciitilde3 tokens.
We use this observation to create a \textbf{Sliding Window CPT Filter}: we apply the same filtering mechanism as before, 
but instead of applying it to the entire prompt, it is applied to the different subsets of the input prompt, after converted to tokens.
An example can be seen in Figure \ref{fig:rolling_cpt}, where a sliding window of size 5 tokens is applied to a mixed text.

\section{Conclusion}
\raggedright In this work, we introduced \textbf{CPT-Filtering}, a simple, model-agnostic, and computationally-negligible guardrail to defend against jailbreak attacks that use encoded and 
ciphered text. We have shown that the inherent nature of BPE tokenizers causes them to represent these out-of-distribution inputs inefficiently, a behavior that 
can be reliably detected by measuring the average characters per token. 

We have demonstrated that:
\begin{itemize}
    \item BPE tokenizers can be used to distinguish obfuscated prompts from regular language with \textbf{near-perfect results}.
    \item Distinction can be applied to very short strings, thus allowing the use of sliding-windows to identify mixed inputs.
    \item Monolingual BPE tokenizers can be used to distinguish obfuscated texts from different languages.
    \item BPE tokenizers are better in obfuscation detection than standard perplexity filtering, both in performance and computation costs.
\end{itemize}

As we have established, our findings support two critical security applications: 
\begin{enumerate}
    \item In online systems, CPT-Filtering can act as an efficient, real-time guardrail to filter potentially harmful prompts before they reach the LLM. 
    \item For model development, the method provides a powerful tool for data curation, helping to identify and remove poisoned examples from training and fine-tuning datasets. 
\end{enumerate}
We also present our dataset of \textasciitilde120k prompts, both obfuscated and natural, to support further research of the community. 
\section{Acknowledgments}
The authors would like to thank Dor Ringel for his great assistance in reviewing this research.

\bibliographystyle{plainnat}
\bibliography{bib}

\clearpage
\appendix
\section{Appendix}

\subsection{Leetspeak}
Leetspeak is a form of coded language where letters are replaced with numbers or symbols, and is commonly used in online communities.
Since Leetspeak in not a standard encoding method, prompts may have different encoding styles. In this paper, when we refer to Leetspeak, we use the
mapping shown in Table~\ref{tab:leetspeak}.

\begin{table}[htb]
    \centering
    \begin{tabular}{c c}
        \hline
        \textbf{Letters} & \textbf{Leetspeak} \\
        \hline
        A, a & 4 \\
        B, b & 8 \\
        E, e & 3 \\
        G, g & 6 \\
        I, i, L, l & 1 \\
        O, o & 0 \\
        S, s & 5 \\
		T, t & 7 \\
        \hline
    \end{tabular}
    \caption{Leetspeak mapping}
    \label{tab:leetspeak}
\end{table}

\subsection{T-Test of the Separation Between the Original and Obfuscated Prompts}
We display here (Table~\ref{tab:t_test_full}) the results of t-statistics and p-values for all comparison between the original and obfuscated prompts, 
segregated by the type of obfuscation.

\begin{table*}[htb]
    \centering
    \footnotesize
    \begin{tabular}{l c c c c c c c c c c}
        \hline
		 & \multicolumn{2}{c}{Reversed} & \multicolumn{2}{c}{Base64} & \multicolumn{2}{c}{Caesar} & \multicolumn{2}{c}{Leetspeak} & \multicolumn{2}{c}{Binary} \\
        \textbf{Tokenizer} & \textbf{t-stat} & \textbf{p-val} & \textbf{t-stat} & \textbf{p-val} & \textbf{t-stat} & \textbf{p-val} & \textbf{t-stat} & \textbf{p-val} & \textbf{t-stat} & \textbf{p-val} \\
        \hline
        Qwen-3 & 406 & $< 10^{-5}$ & 561 & $< 10^{-5}$ & 473 & $< 10^{-5}$ & 603 & $< 10^{-5}$ & 630 & $< 10^{-5}$  \\
        GPT-4o & 411 & $< 10^{-5}$ & 585 & $< 10^{-5}$ & 510 & $< 10^{-5}$ & 595 & $< 10^{-5}$ & 459  & $< 10^{-5}$ \\
		Llama-3.2 & 441 & $< 10^{-5}$ & 604 & $< 10^{-5}$ & 513 & $< 10^{-5}$ & 592 & $< 10^{-5}$ & 455 & $< 10^{-5}$ \\
        Gemma-3 & 381 & $< 10^{-5}$ & 528 & $< 10^{-5}$ & 454 & $< 10^{-5}$ & 590 & $< 10^{-5}$ & 616 & $< 10^{-5}$ \\
        \hline
    \end{tabular}
    \caption{t-test of the separation between the original and all obfuscated prompts for each tokenizer.
	        Every segment has 19,816 examples.}
    \label{tab:t_test_full}
\end{table*}

\subsection{Additional Figures}

\begin{figure*}[tb]
    \centering
    \includegraphics[width=0.9\textwidth]{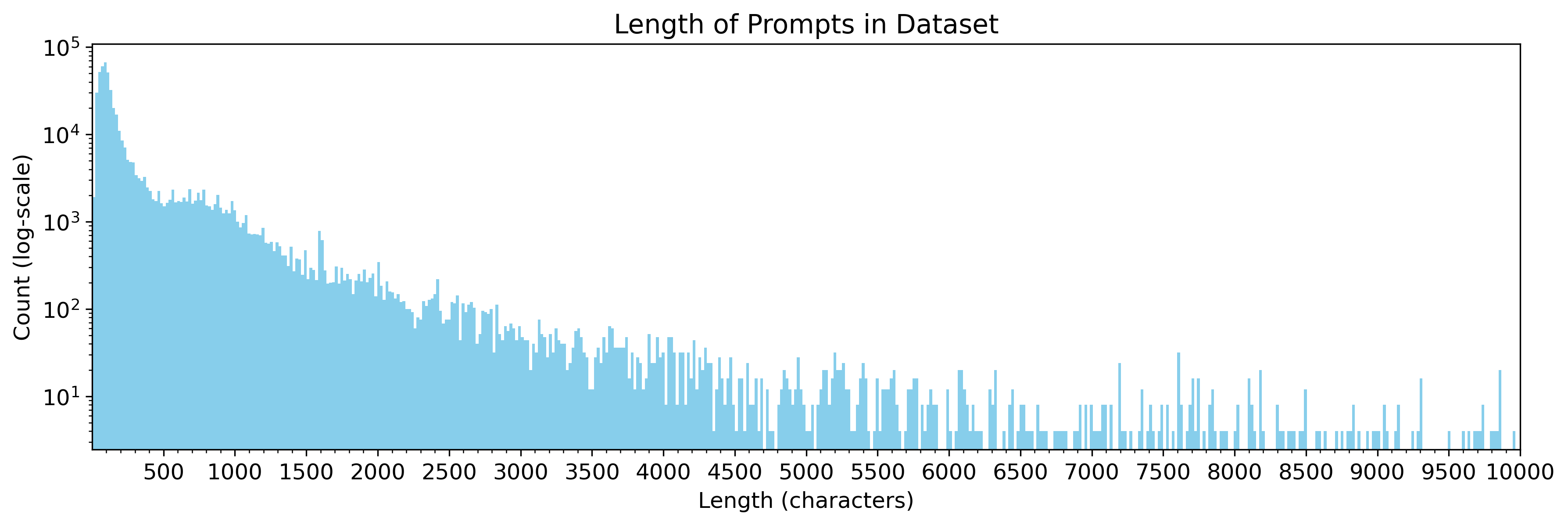} 
    \caption{Length of prompts in the dataset (in characters)}
    \label{fig:dataset}
\end{figure*}

\begin{figure*}[tb]
    \centering
    \includegraphics[width=\textwidth]{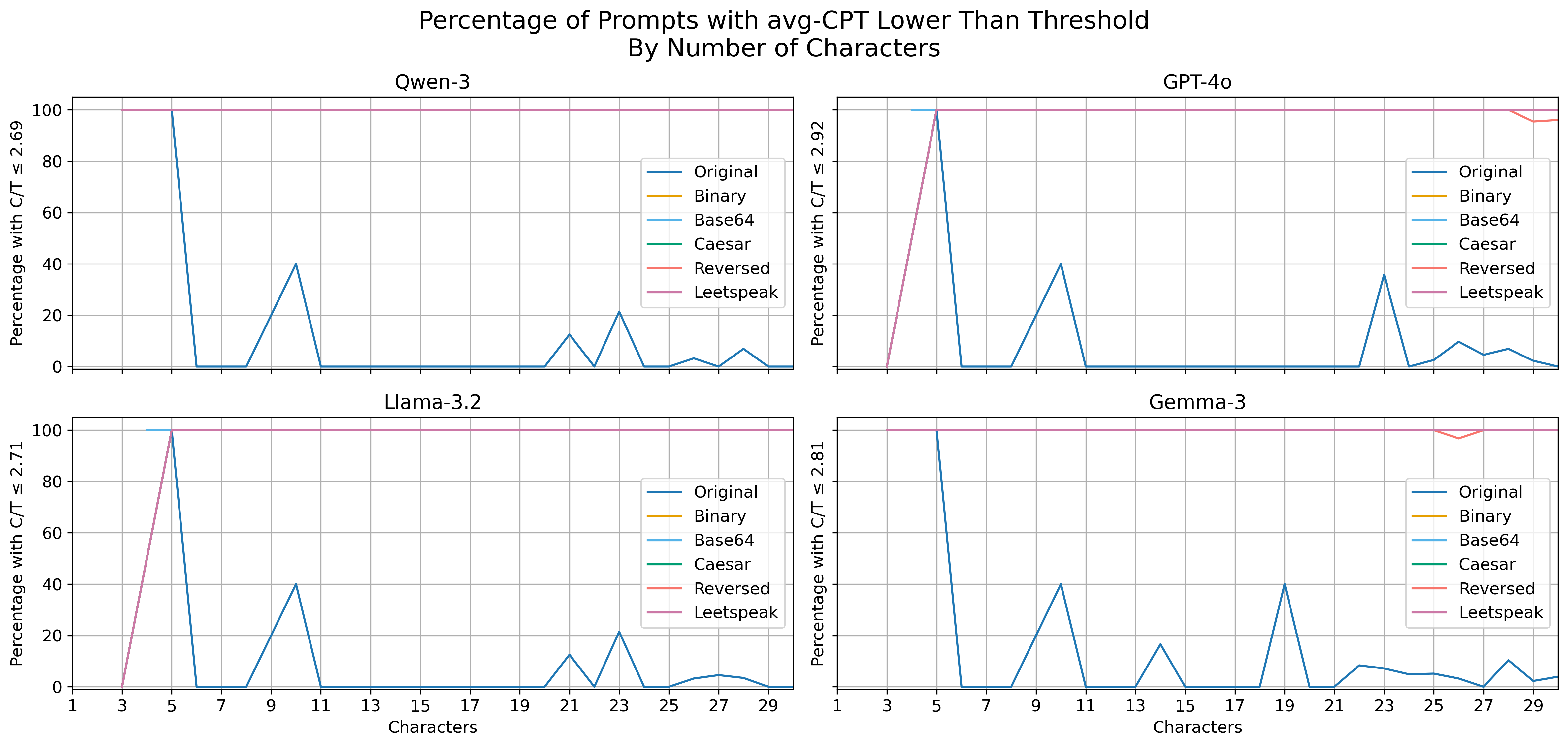} 
    \caption{\raggedright Percentage of prompts marked as ``obfuscated-prompts'' in each category, as a function of prompt length characters}
    \label{fig:short_chars}
\end{figure*}

\begin{figure*}[tb]
    \centering
    \includegraphics[width=\textwidth]{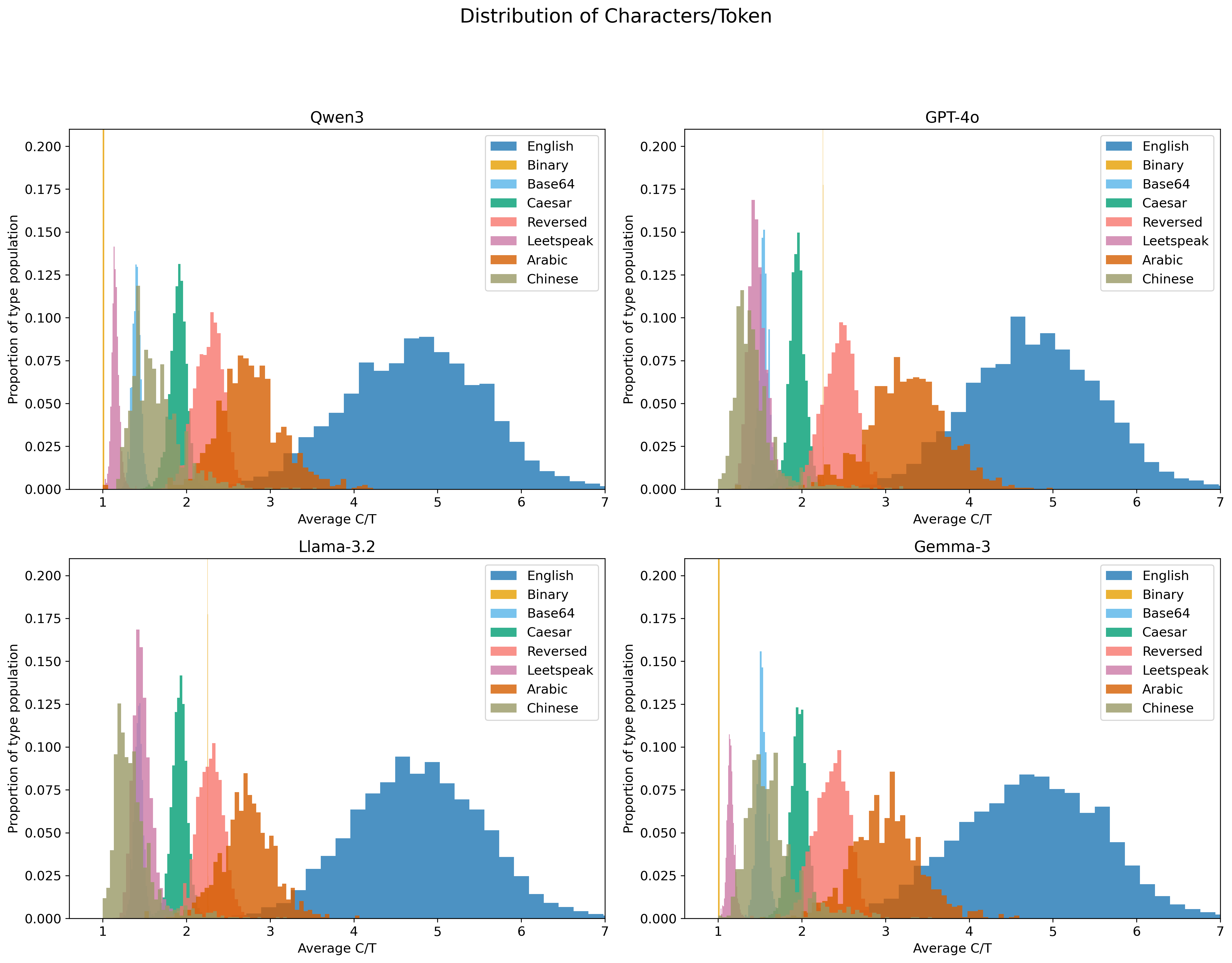} 
    \caption{\raggedright The average number of characters-per-token for the four models checked, with Chinese and Arabic languages included.}
    \label{fig:langs_no_good}
\end{figure*}

\clearpage

\section{Code Appendix}
For reproducibility, we provide the code used to create the dataset used in the main text.
Please note it requires signing up for an account on the Hugging Face platform, and accepting the terms of service of the datasets (all are licensed under MIT License, yet some require pre-approval on the platform).
The code contains the exact revisions used to create the dataset.

\begin{lstlisting}[language=Bash, caption={Install Dependencies}, label={lst:install_dependencies}]
pip install pandas datasets
\end{lstlisting}

\begin{lstlisting}[language=Python, caption={Dataset Creation Code}, label={lst:dataset_creation}, float=*]
import base64
import pandas as pd
from datasets import load_dataset


def b64_encode(text: str) -> str:
    return base64.b64encode(text.encode()).decode()

def reversed_text(text: str) -> str:
    return text[::-1]       

def caesar_encode(text: str, shift: int = 3) -> str:
    result = ""
    for char in text:
        if char.isalpha():
            # Determine the case and base ASCII value
            ascii_base = ord('A') if char.isupper() else ord('a')
            # Apply shift and wrap around alphabet using modulo
            shifted = (ord(char) - ascii_base + shift) % 26
            result += chr(ascii_base + shifted)
        else:
            result += char
    return result

def leetspeak(text: str) -> str:
    return text.translate(str.maketrans('abegilostABEGILOST', '483611057483611057'))

def text_to_binary(text):
    binary = ' '.join(format(ord(char), '08b') for char in text)
    return binary

if __name__ == "__main__":
    df = pd.concat([
        load_dataset("SoftAge-AI/prompt-eng_dataset", revision='df8d7daa0affe6d1ffe79dcdebad5cc3bbf38ce5')['train'].to_pandas()[['Prompt']].rename({'Prompt': 'prompt'}, axis=1),
        load_dataset("Aiden07/dota2_instruct_prompt", revision='ac11696ad02b6e21c1412e7dad7b05e858806ff6')['train'].to_pandas()[['instruction']].rename({'instruction': 'prompt'}, axis=1),
        load_dataset("hassanjbara/ghostbuster-prompts", revision='22dbacbad1a3060946ee66fa2e1447a17adf0ac4')['train'].to_pandas()[['text']].rename({'text': 'prompt'}, axis=1),
        load_dataset("k-mktr/llm_eval_prompts", revision='28ebecb5043f06d9da3abaf0a03a204404c37c98')['test'].to_pandas()[['prompt']],
        load_dataset("facebook/cyberseceval3-visual-prompt-injection", revision='7933662024dc994be4ab90d520ab712e5765b655')['test'].to_pandas()[['user_input_text']].rename({'user_input_text': 'prompt'}, axis=1),
        load_dataset("HacksHaven/science-on-a-sphere-prompt-completions", revision='95daacaf5415856e6768331937d3e779f308e7d3')['train'].to_pandas()[['prompt']],
        load_dataset("grossjct/ethical_decision_making_prompts", revision='b7b2bbf6ff3a23f6411bf718b4462d5e1758a654')['train'].to_pandas()[['prompt']],
    ], 
    ignore_index=True)

    df = df[df['prompt'].apply(lambda x: bool(str(x).strip()))].reset_index(drop=True)
    df = pd.concat([
        pd.DataFrame({'prompt': df['prompt'], 'type': 'Original'}),
        pd.DataFrame({'prompt': df['prompt'].apply(b64_encode), 'type': 'Base64'}),
        pd.DataFrame({'prompt': df['prompt'].apply(reversed_text), 'type': 'Reversed'}),
        pd.DataFrame({'prompt': df['prompt'].apply(caesar_encode), 'type': 'Caesar'}),
        pd.DataFrame({'prompt': df['prompt'].apply(leetspeak), 'type': 'Leetspeak'}),
        pd.DataFrame({'prompt': df['prompt'].apply(text_to_binary), 'type': 'Binary'}),
    ], ignore_index=True)

    df.to_csv('dataset.csv', index=False)
\end{lstlisting}

\end{document}